\documentstyle[ijbc,12pt]{article}
\input{epsf}
\begin{document}

\title{\vspace*{-4cm} Heteroclinic Contours in Neural Ensembles and
the Winnerless Competition Principle}

\author{Valentin S. Afraimovich$^{1}$, Mikhail I. Rabinovich$^{2}$,
Pablo Varona$^{2,3}$\\ \\
\small $^{1}$Instituto de Investigaci\'on en Comunicaci\'on
  ~\'Optica, UASLP.\\ \small A. Obreg\'on 64. 78000 San Luis
Potos\'{\i}, SLP, M\'exico.\\ 
\small $^{2}$Institute for Nonlinear Science.University of California San Diego. \\ \small 9500 Gilman Dr. La Jolla. CA
92093-0402.\\ 
\small $^{3}$GNB. Dpto. de Ingenier\'{\i}a Inform\'atica. \\ \small Universidad Aut\'onoma de Madrid. 28049 Madrid, Spain.
}

\date{Dec 26, 2002}

\maketitle

\begin{abstract}  
  The ability of nonlinear dynamical systems to process incoming
information is a key problem of many fundamental and applied
sciences. Information processing by computation with attractors
(steady states, limit cycles and strange attractors) has been a
subject of many publications. In this paper we discuss a new direction
in information dynamics based on neurophysiological experiments that
can be applied for the explanation and prediction of many phenomena in
living biological systems and for the design of new paradigms in
neural computation.  This new concept is the Winnerless Competition
(WLC) principle. The main point of this principle is the
transformation of the incoming identity or spatial inputs into
identity-temporal output based on the intrinsic switching dynamics of
the neural system. In the presence of stimuli the sequence of the
switching, whose geometrical image in the phase space is a
heteroclinic contour, uniquely depends on the incoming
information. The key problem in the realization of the WLC principle
is the robustness against noise and, simultaneously, the sensitivity
of the switching to the incoming input. In this paper we prove two
theorems about the stability of the sequential switching and give
several examples of WLC networks that illustrate the coexistence of
sensitivity and robustness.

\vspace*{1cm}
\noindent{\bf Paper in press for 'International Journal of Bifurcation
  and Chaos' (2004), Vol. 14 (4).}

\end{abstract}

\newpage

 \section{Introduction}

Computing with a dynamical system implies that this system
changes its behavior depending on the quality and quantity of the
incoming information.  This is an enormous field and we will
concentrate here only on the concept of Winnerless Competition (WLC)
that, as we think, is a general principle for information processing by
dynamical systems.

Information processing with WLC dynamics is a new area for theoretical
study. However, WLC
itself has already been observed in many well known experiments in
hydrodynamics~\cite{busse}, population biology~\cite{May} and laser
dynamics~\cite{Roy}. For example, the convective roll patterns in a
rotating plane layer demonstrate a sequential changing of direction as
a result of the competition between different roll orientations (when
the rotation rate is large enough, e.g.  Kuppers-Lortz
instability~\cite{Kuppers}).  For a large Prandtle number the critical
angle is close to $60^o$ (for the steady state rolls it is $120^o$). Due to
the appearance of new rolls that are also unstable, no pattern becomes
a winner and, as a result of the competition, they switch
sequentially. Small non-Boussinesq effects are able to make such
sequence periodic~\cite{misha}. 

The discussed process can be described
by Lottka-Volterra equations that are well known in population biology
and can explain the competition between three spices~\cite{May}:
\begin{eqnarray}
\dot{a}_1 & = & a_1[1-(a_1+\rho_{12}a_2+\rho_{13}a_3)] \nonumber \\
\dot{a}_2 & = & a_2 [1-(a_2+\rho_{21}a_1+\rho_{23}a_3)] \nonumber \\
\dot{a}_3 & = & a_3 [1-(a_3+\rho_{31}a_1+\rho_{32}a_3)] \nonumber 
\end{eqnarray}

In biology the variable $a_i(t)$ is the number of individuals in the
$i$--th population at time $t$, and $\rho_{ij}$ are competition
coefficients measuring the extent to which the $j$th species affects the
growth rate of the $i$--th species.  In convection, the variables $a_i$ are the
intensities of the competitive modes: $a_i=|c_i|^2$, and we suppose that
the vertical component of the velocity field is (in the limit of small
amplitudes) $u_x=f(z)\sum_{j=1}^{3}c_j(t)exp\{i{\bf k}_j\cdot \bf{r}\}$,
where $z$ is the component of the position vector ${\bf r}$ in the
vertical direction and ${\bf k}_j$ are the wave vectors~\cite{busse,misha}.

The non-symmetry of the coefficients $\rho_{ij}$, for example
$\rho_{12}=\rho_{23}=\rho_{31}\equiv \rho_{+}>1$,
$\rho_{21}=\rho_{32}=\rho_{13}\equiv \rho_{-}<1$, guarantees the
WLC behavior of the discussed dynamical system. The mathematical image
of such behavior is a heteroclinic contour in the phase space $a_1(t)$,
$a_2(t)$, $a_3(t)$ (see Fig.~\ref{Fig1}). Results related to symmetric
cases have been reporter earlier \cite{ashwin1} and \cite{ashwin2}.


The questions that we are going to discuss below are: (i) what are the
conditions for the robustness of the WLC, i.e. the topological
similarity of the perturbed and original heteroclinic contour; and
(ii) how subsystems with WLC behavior interact with each other. 

The paper is organized in the following way. First we describe a class
of models that use the WLC principle for the representation and
processing of incoming information (Sec.~\ref{models}). Then we
discuss the existence and stability of the heteroclinic contour
(Sec.~\ref{existence}), and the robustness, i.e. birth of a stable
limit cycle in the vicinity of the destroyed heteroclinic loop in a
perturbed system (Sec.~\ref{stable}) with some examples from computer
modeling. Finally we discuss several WLC strategies used by
living neural systems to perform complex information processing
(Sec.~\ref{realn}).

\section{The Models}
\label{models}

The activity of many different neural
networks~\cite{mishaprl,clione,abeles,cohen}, can be described
qualitatively with the following dynamics:

\begin{eqnarray}
\dot{a}_i & = & a_i(\sigma(\mbox{\boldmath $H$},\mbox{\boldmath
  $S$})-\sum_{j=1}^{N}\rho_{ij}a_j+H_i(t))+S_i(t)
\label{eq2}
\end{eqnarray}

\noindent where $a_i>0$ represents the instantaneous spiking rate of
the principal neurons (PNs) that are making the computation,
$\rho_{ij}$, represents the strength of inhibition in $i$ by $j$,
$H_i(t)$ represents the action from other neural ensembles, and
$S_i(t)$ represents the stimuli from the sensors.  In many neural
networks, the inhibition among PNs is the result of the action of
inhibitory local neurons (LNs). Usually LNs also receive an external
input and because of this $\rho_{ij}$ can depend on the stimuli.

The dynamical system (\ref{eq2}) in the case $\sigma=1$,
$\mbox{\boldmath $H$}(t)=\mbox{\boldmath $S$}(t)=0$ is the 
Lottka-Volterra model.  The dynamics of the system is well known when
the matrix $\rho_{ij}$ is symmetric ($\rho_{ij}=\rho_{ji}$). In this
case the autonomous system has a global Lyapunov
function~\cite{cohen,hopfield} and every trajectory approaches one of
the numerous possible equilibrium points. For example, if the
inhibitory connections are identical, $\rho_{ij}=\rho$, $\rho_{ii}=1$,
this system has only one global attractor, e.g.
$a_{i}=a_{0}=1/[1+\rho\,(N-1)]$ for $\rho<1$, and $N$ attractors:
$a_{i}=a_{0}=1$, $a_{j\neq i}=0$ if $\rho >1$.  No other attractors,
e.g. limit cycles, or strange attractors are present in the
system. The situation is much more complex and interesting when the
inhibition is non-symmetric: $\rho_{ij} \neq \rho_{ji}$. A detailed
analysis is only possible in the case $N=3$ (see
references~\cite{afraa1,chi,mishaprl}). When $\rho_{ij}>1, \rho_{ji}<1$
there exists a heteroclinic contour in the phase space of the system
that consists of saddle points and one-dimensional separatrices
connecting them. In some regions of the parameter space $\{\rho_{ij} \}$,
such heteroclinic contour (or limit cycle in its vicinity) is a global
attractor. If $\rho_{ij}$ depend on the stimulus, e.g. as a result of
a learning mechanism, the system (\ref{eq2}) can generate different
heteroclinic contours for different stimuli~\cite{mishaprl}.

 Suppose the matrix

\begin{center}
\begin{math}(\rho_{ij})=\left( \begin{array}{ccc}1&\alpha_{1}&\beta_{1}\\
                                    \beta_{2} & 1         & \alpha_{2}
\\
                                    \alpha_{3} & \beta_{3} &
1\end{array} \right)
                                    \end{math}
\end{center}

\noindent and $0<\alpha_{i}<1<\beta_{i}$ and
$\kappa_{i}=(\beta_{i}-1)/(1-\alpha_{i})$.  Then the heteroclinic
contour is a global attractor if
$\kappa_{1}\cdot\kappa_{2}\cdot\kappa_{3}>1$, and the nontrivial fixed
point $A(a_{1}^{0},a_{2}^{0},a_{3}^{0})$ is a saddle point. If
$\kappa_{1}\cdot\kappa_{2}\cdot\kappa_{3}=1$, this fixed point becomes
neutrally stable and there exists a family of neutrally stable
periodic solutions in the phase space. When
$\kappa_{1}\cdot\kappa_{2}\cdot\kappa_{3}<1$, $A$ becomes a global
attractor. The heteroclinic orbit exists but looses its stability. It
is important to emphasize that in the case
$\kappa_{1}\cdot\kappa_{2}\cdot\kappa_{3}>1$ a small perturbation is
able to destroy the heteroclinic orbit and then a stable limit cycle
appears in its vicinity. This limit cycle is characterized by a finite
time period of switching among different states, in contrast with the
infinite time of motion along the heteroclinic loop.

When $N>3$ the dynamics of system (\ref{eq2}) can be very complex and even
chaotic~\cite{clione}.  Here we are interested in the existence and
stability of the heteroclinic contours, which are the mathematical
image of the winnerless competition behavior. Such orbits may exist
only in the nonsymmetric case e.g.  $\rho_{ij} \neq \rho_{ji}$, when
saddle points (in the heteroclinic contours) satisfy several conditions.

\section{Existence and Stability of the Heteroclinic Contour}
\label{existence}

 In this section we consider the canonical Lottka-Volterra model

\begin{eqnarray}
\dot{a}_i & = & a_i[1-(a_i+\sum_{i \neq j}^{N}\rho_{ij}a_j)]\label{eq3},
\end{eqnarray}

\noindent  and derive conditions of existence and stability of heteroclinic 
  contours.

\subsection{Necessary Conditions}

\subsubsection{"Codimension one" saddle points}

A heteroclinic contour consists of finitely many saddle equilibria and
finitely many heteroclinic orbits connecting these equilibria. Let's
denote by $A_1$ the equilibrium point $(1,0,0,...,0)$, by $A_2$ the
point $(0,1,...0)$, and by $A_N$ the point $(0,0,...,1)$. For the sake
of simplicity we assume that there is a heteroclinic orbit $\Gamma_{i i+1}$
connecting the points $A_i$ and $A_{i+1}$, $i=1,...N$ and $A_{N+1}\equiv
A_1$. (If not, we can always apply a change of variables in the
form of a permutation). 
The contour can serve as an
attracting set if every point $A_i$ has only one unstable direction. 
By direct verification it can be shown that $A_i$
satisfies this assumption provided that:

\begin{eqnarray}
\rho_{k i} >1, k \neq i+1\label{eq7},
\end{eqnarray}

\noindent and

\begin{eqnarray}
\rho_{i+1 i} <1\label{eq8}.
\end{eqnarray}

\noindent (Here $i+1=1$ if $i=N$).

Moreover, if (\ref{eq7}) and (\ref{eq8}) are satisfied then the unstable
direction at the point $A_i$ is parallel (at that point) to the
ort $(0...0 1 0...0)$, where 1 corresponds to the $i$--th coordinate. An
intersection of hyperplanes $P_{2i}=\bigcap_{j=1,j\neq i, i+1}^N\{a_j =
0\}$ is a two-dimensional invariant manifold containing points $A_i$
and $A_{i+1}$ such that $A_i$ is a saddle point on $P_{2i}$ and
$A_{i+1}$ is a stable node on $P_{2i}$. The system (\ref{eq3}) on $P_{2i}$
has the form:

\begin{eqnarray}
\dot{a}_i & = & a_i[1-(a_i+\rho_{i i+1}a_{i+1})]\\
\dot{a}_{i+1} & = & a_{i+1}[1-(a_{i+1}+\rho_{i+1 i}a_{i})]\label{eq9}
\end{eqnarray}

\noindent and, from (\ref{eq7}) and (\ref{eq8}), one has $\rho_{i i+1}>1$,
$\rho_{i+1 i}<1$.

This implies that there are no equilibrium points in the region
$a_i>0$, $a_{i+1}>0$, and since $\dot a_{i+1}<0$ if $a_{i+1} \gg 1$
then it is simple to see that the separatrix, say $\Gamma_i$ of the
saddle point $A_i$ must go to the attractor $A_{i+1}$, i.e. there is a
heteroclinic connection between $A_i$ and $A_{i+1}$ on the plane
$P_{2i}$ (For the case N=3 see the proof in~\cite{Waltman}).

\subsubsection{Leading directions}

The point $A_i$ on $P_{2i}$ is a stable node with characteristic
numbers $\lambda_1=-1$ and $\lambda_2=1-\rho_{i i+1}$. The leading
direction at $A_{i+1}$ is determined  by the absolute values of
$\lambda_1$ and $\lambda_2$: if $\lambda_1 > \lambda_2$ then the
leading direction is parallel to the $a_{i+1}$-axis, and if $\lambda_1
< \lambda_2$ then the leading direction is transversal to the
$a_{i+1}$-axis on $P_{2i}$. We assume that the last inequality
 holds, i.e. 

\begin{equation}
\rho_{i i+1} < 2,
\label{smaller}
\end{equation}
then the majority of orbits, (including $\Gamma_i$) go to $A_{i+1}$
following a direction $\vec{l}=(1,-\rho_{i+1 i}/(2-\rho_{i i+1}))$
transversal to the $a_{i+1}$-axis.

The vector $\vec{l}$ on $P_{2i}$ can be embedded into the hyperplane 
$H_i : \{a_{i+2}=0\}$ as $\vec{L}=(0 0 ... 1, -\rho_{i+1 i}/(2-\rho_{i i+1}), 0...0)$, $1$ on the $i$--th place, and one can ask if the direction
$\vec{L}$ is the leading direction for the node $A_{i+1}$ on this
hyperplane. To see sufficient conditions for that assumption we have
to take into account that the characteristic numbers at point $A_{i+1}$ of
the system (\ref{eq3}) restricted to the hyperplane 
$\{a_{i+2}=0\}$ are $1-\rho_{i
i+1},...,1-\rho_{i-1 i+1}, -1, 1-\rho_{i+2 i+1},..., 1-\rho_{Ni+1}$
(they are all negative because of (\ref{eq7})). Hence, if

\begin{eqnarray}
 \rho_{k i+1} > \rho_{i i+1}, k \neq i\label{eq10},
\end{eqnarray}

\noindent then $1-\rho_{i i+1}$ is the characteristic value closest to
zero and $\vec{L}$ is the leading direction at
$A_{i+1}$ on $H_i$. 
We assume that (\ref{eq10}) is
satisfied. This condition is not necessary for the validity of the
results below, but it essentially simplifies the description of the 
results and calculations.

\subsubsection{Saddle values}

The point $A_{i}$ is a saddle point on $P_{2i}$. One can write a map
from a transversal to the stable separatrix into a transversal to the
unstable separatrix along the orbits going through a neighborhood of
$A_i$ (see for instance \cite{shilnikov}). In suitable coordinates $(\xi,\eta)$ it
has the form:

\begin{eqnarray}
\xi = c \eta^{\nu_{i}}\label{eq11}
\end{eqnarray}

\noindent where $\eta$ is a deviation from the stable manifold,
$\xi$ is a deviation from the unstable one, $c$ is a constant and

\begin{eqnarray}
\nu_i = - \frac{1- \rho_{i i+1}}{1- \rho_{i+1 i}} \equiv \frac{\rho_{i
i+1} -1 }{1- \rho_{i+1 i}}\label{eq12}
\end{eqnarray}
is the ``saddle value'' (\cite{shilnikov}). If $\nu_i >1$ then the map (\ref{eq11})
is a local contraction and $P_i$ is a dissipative saddle. If $\nu_i <
1$ then (\ref{eq11}) is a local expansion.

\subsubsection{Stability of the heteroclinic contour}

The following result tells us that the contour $\Gamma=\bigcup_{i=1}^N
\Gamma_i \cup A_i$ can be an attractor.

\newtheorem{theorem}{Theorem}

\begin{theorem}
 Assume that
conditions (\ref{eq7}), (\ref{eq8}), (\ref{smaller}), (\ref{eq10}) are satisfied
and

\begin{eqnarray}
\nu = \prod_{i=1}^N \frac{\rho_{i i+1} -1}{1- \rho_{i+1 i}} >
1\label{eq13}
\end{eqnarray}

\noindent (here $i+1=1$ if $i=N$). Then there is a neighborhood $U$
of the contour $\Gamma$ such that for any initial condition
$a^0=(a_1^0,...,a_N^0)$ in $U$ with $a^0_i>0$, one has dist$(a(t), \Gamma)
\rightarrow 0$ as $t \rightarrow \infty$ where $a(t)$ is the orbit going
through $a^0$.
\end{theorem}

\subsection{Proof of Theorem 1}
\label{Proof}

The proof of the theorem is based on the construction of the
Poincar\'e map along orbits in a neighborhood of the contour
$\Gamma$. Let $W_i^s$ ($W_i^u$) be a stable (unstable) manifold of the
point $A_i$ and $P_i$ ($Q_i$) be a point on the heteroclinic orbit
$\Gamma_{i-1}$ ($\Gamma_{i}$) in a small neighborhood of 
$A_i$--see Fig \ref{trayectof}.


\subsubsection{Local map}

Let $S_{P_i}$ ($S_{Q_i}$) be a piece of a transversal to $\Gamma_{i-1}$
(to $\Gamma_{i}$) hyperplane going through $P_i$ (through
$Q_i$). Without loss of generality, we may assume that $S_{P_i}$ is a
piece of a hyperplane, parallel to the hyperplanes ${a_{i+1=0}}$. A
local map $f_i:S_{P_i} \rightarrow S_{Q_i}$ along orbits in a neighborhood
of $A_i$ is well defined. In suitable coordinates it has the form 
\cite{shilnikov}.

\begin{eqnarray}
\xi_i = c_i \eta_i^{\nu_{i}}, \chi_i=\varphi_i(y_i,\eta_i)\label{eq14}
\end{eqnarray}

\noindent where $\eta_i \in \Re$ is a coordinate on $S_{P_i}$,
``parallel'' to $W_i^u$, $y \in \Re^{N-2}$ is a vector-coordinate
transversal to the $\eta_i$-axis on $S_{P_i}$, $\xi_i$ is a coordinate
on $S_{Q_i}$, ``parallel'' to the leading direction on $W_i^S$ at
$A_i$, $y_i$ is a vector-coordinate transversal to the $\xi_i$-axis on
$S_{Q_i}$. Moreover,

\begin{eqnarray}
|\frac{\partial \varphi_i}{\partial y_i}| \leq \bar{c}_i
 \eta_i^{\beta_i}, |\frac{\partial \varphi_i}{\partial \eta_i}| \leq \bar{c}_i
 |y_i| \cdot |\eta_i|^{\beta_i-1}\label{eq15}
\end{eqnarray}

\noindent where $\bar{c}_i>0$ is a constant and

\begin{eqnarray}
 \beta_i > \nu_i \label{eq15b}
\end{eqnarray}

\subsubsection{Global map}

The heteroclinic orbit $\Gamma_i$ has a piece joining the points $Q_i$
and $P_{i+1}$. Thus, the global map $F_i:\tilde{S}_{Q_i} \rightarrow
S_{P_{i+1}}$ along orbits in a neighborhood of this piece is
well-defined where $\tilde{S}_{Q_i} \subset S_{Q_i}$ is a small
neighborhood of the point $Q_i$ on $S_{Q_i}$. This map is a
diffeomorphism and has the form

\begin{eqnarray}
\eta_{i+1} & = & a_{i 1}\xi_i + a_{i 2}\xi_i + ... \nonumber \\
y_{i+1} & = & b_{i 0} + b_{i 1}\xi_i + b_{i 2}\xi_i + ... \label{eq16}
\quad ,
\end{eqnarray}
where the dots denote nonlinear terms. The orbit $\Gamma_i$ belongs to
the intersection of invariant hyperplanes $\{a_j=0\}$, $j \neq i$, $j
\neq i+1$. Therefore the hyperplane $\{\xi_i=0\}$ on $S_{Q_i}$ is mapped by
$F_i$ into the hyperplane $\{\eta_{i+1}=0\}$ on $S_{P_{i+1}}$ which means
that $a_{i2}=0$ in (\ref{eq16}), and

\begin{eqnarray}
a_{i1} \neq 0\label{eq17}
\end{eqnarray}

Hence (\ref{eq16}) has the form:

\begin{eqnarray}
\eta_{i+1} & = & a_{i}\xi_i + ... \nonumber \\
y_{i+1} & = & b_{i 0} + b_{i 1}\xi_i + b_{i 2}\xi_i + ... \quad , \label{eq18}
\end{eqnarray}

$a_{i1}=:a_i$.

\subsubsection{Poincar\'e map}

We may construct a Poincar\'e map $F=S_{P_1} \rightarrow S_{P_1}$
as the superposition of maps $f_i$, $F_i$, i.e. $F$ has the form
$F=F_N \cdot f_N \cdot ... \cdot F_2 \cdot f_2 \cdot F_1 \cdot f_1$.

The map $F_i \circ f_i$ has the form 

\begin{eqnarray}
\eta_{i+1} & = & a_{i} c_i \eta_i^{\nu_i} + ... \nonumber \\
y_{i+1} & = & b_{i 0} + b_{i 1} c_i \eta_i^{\nu_i} + b_{i 2} \varphi_i(y_i,\eta_i) + ... \label{eq19}
\end{eqnarray}

So, in this approximation the form of the map $F$ along the
$\eta$-coordinates is independent of the $y$-coordinates , and one may
consider the one-dimensional approximation

\begin{eqnarray}
\eta_{i+1} & = & a_{i} c_i \eta_i^{\nu_i}, i=1,...,N \label{eq20} \\
\eta_{N+1} & := & \bar{\eta}_1 \equiv \tilde{F}(\eta_1)\nonumber
\end{eqnarray}

It is simple to see that the map $\tilde{F}:\eta_{1} \rightarrow
\bar{\eta}_{1}$ is a contraction provided that condition (\ref{eq13}) is
satisfied. Indeed, it follows from (\ref{eq20}) that $\bar{\eta_1} = A
\eta_1^\nu$, where $A$ is a constant and
$\nu=\prod_{i=1}^N\nu_i$. Hence if $\nu>1$, then $\partial \bar{\eta_1} /
\partial \eta_1 <1$ if $\eta_1$ is small enough, the map (\ref{eq20}) is a
contraction and $\eta_1=0$ is an attracting fixed point (corresponding
to the contour $\Gamma$).

Consider now the map $F_i \circ f_i$. The differential

\begin{eqnarray}
D F_i\circ f_i = \left( \begin{array}{ll}
                         a_i c_i \nu_i \eta_i^{\nu_i -1} +... & 0+... \\
                         b_{i 1} c_i \nu_i \eta_i^{\nu_i -1} + b_{i 2}
                         \frac{\partial \varphi_i}{\partial \eta_i} +
                         ... & \frac{\partial \varphi_i}{\partial y_i}+...
			\end{array} \right)
\end{eqnarray}

\noindent and because of (\ref{eq15}), (\ref{eq15b}),

\begin{eqnarray}
\| D F_i \circ f_i \| \leq B_i \eta_i^{\nu_i -1} \label{eq21}
\end{eqnarray}

\noindent where $B_i$ is a constant. Since $D F = \prod_{i=1}^N D(F_i
\circ f_i)$,  then 

\begin{eqnarray}
\| D F \| \leq B \prod_{i=1}^N \eta_i^{\nu_i -1} \label{eq22}
\end{eqnarray}

\noindent where $B = \prod_{i=1}^N B_i$. We estimate now
$\prod_{i=1}^N \eta_i^{\nu_i -1}$ by using (\ref{eq19}). In the
expression $\eta_N^{\nu_N-1} \cdot \eta_{N-1}^{\nu_{N-1}-1} \cdot
... \cdot \eta_2^{\nu_2-1} \cdot \eta_1^{\nu_1-1}$ let us estimate
first $\eta_2^{\nu_2-1} \cdot \eta_1^{\nu_1-1}$. Since  $\eta_2=a_1
c_1 \eta_1^{\nu_1}+...$, then $\eta_2^{\nu_2-1} \cdot \eta_1^{\nu_1-1}
= (a_1 c_1)^{\nu_2-1} \cdot (\eta_1^{\nu_1}+...)^{\nu_2-1} \cdot
\eta_1^{\nu_1 -1} \leq {\mbox const} \cdot \eta_1^{\nu_1 \nu_2 -1}$. 

Now $\eta_3^{\nu_3-1} \cdot \eta_2^{\nu_2-1} \cdot \eta_1^{\nu_1-1}
= (a_2 c_2)^{\nu_3-1} \cdot (\eta_2^{\nu_2}+...)^{\nu_3-1} \cdot
\eta_1^{\nu_1 -1} \leq const \cdot \eta_2^{\nu_3 \nu_2 -1} \cdot
\eta_1^{\nu_1 -1}= {\mbox const} \cdot (a_1 c_1)^{\nu_3 \nu_2 -1}
(\eta_1^{\nu_1}+...)^{\nu_3 \nu_2 -1} \cdot \eta_1^{\nu_1-1} \leq
{\mbox const} \cdot \eta_1^{\nu_3 \nu_2 \nu_1 -1}$.

Assume (inductively) that  $\eta_k^{\nu_k -1}\cdot ... \cdot \eta_2^{\nu_2
-1} \leq {\mbox const}\cdot \eta_2^{\nu_k ... \nu_2 -1}$, then $\eta_k^{\nu_k -1}\cdot ... \cdot \eta_2^{\nu_2
-1} \cdot \eta_1^{\nu_1 -1} \leq {\mbox const} \cdot (a_1 c_1 \eta_1^{\nu_1} +
...)^{\nu_k ... \nu_2 -1} \cdot \eta_1^{\nu_1 -1} \leq {\mbox const} \cdot
\eta^{\prod_{i=1}^k \nu_i -1}$

\noindent and therefore, 

\begin{eqnarray}
\prod_{i=1}^N \eta_i^{\nu_i -1} \leq C \cdot \eta_1^{\nu - 1}\label{eq23}
\end{eqnarray}

\noindent where $C$ is a constant. Hence,

\begin{eqnarray}
\| D F \| \leq B C \eta_1^{\nu - 1},\label{eq24}
\end{eqnarray}

\noindent i.e. $F$ is a contraction in a neighborhood of the point
$\eta_1=0$, $y_1=0$ corresponding to the contour $\Gamma$. 

This concludes the proof of Theorem 1.


\subsection{Selecting the Contour}
\label{Contour}

  In this subsection we show how the canonical system (\ref{eq3}) can
  be obtained from a general system for a neural network with
inhibitory connections. Suppose that the dynamics of a network of
$N+M$ inhibitory neurons with dynamical variables
$y_i(t)=(y_i^{(1)}(t),...,y_i^{(m)}(t)), i=1,...,N+M$ can be described
in the form of the following system of ODEs:

\begin{eqnarray}
\dot{y}_i & = & F(y_i) - \sum_{j=1}^{N+M} G_{ij}(s)(y_i, y_j) +\tilde S_i(t)\label{eq4}
\end{eqnarray}

\noindent where $F$ is a nonlinear function that describes dynamics of
individual neuron, $G_{ij}(S)(y_i,y_j)$ is a nonlinear operator
describing an inhibitory action of the $j$-th neuron onto the $i$--th
neuron, $S(t)=(S_1(t),...,S_{N+M}(t))$ and
$(\tilde{S}_1(t),...,\tilde{S}_{N+M}(t))$ are the vectors representing
stimuli to the network. We restrict ourselves to a subclass of systems
(\ref{eq4}) represented in the form (\ref{eq2}).  It is known (see
above) that a stimulus acts in two ways: (i) it adds the perturbation
$\tilde{S}(t)$ into (\ref{eq4}) as an external force, and (ii) it
forms the matrix $G_{ij}(S)$.  A simplified model that describes the
firing rate of the neurons can be written in the form
(\ref{eq2})~\cite{mishaprl}, where $\sigma=-1$ when there is no
stimulus, and $\sigma=1$ when the stimulus has a component at neuron
$i$. In the absence of the external force $\tilde{S}(t)$ the system
(\ref{eq3}) is just a subsystem of (\ref{eq2}) for which all
$\sigma=+1$.

We describe now an algorithm to obtain the system (\ref{eq3})
from (\ref{eq2}) (provided that $\tilde{S}(t)=0$).  First of all we
single out only indexes $i$ for which $\sigma_i(S)=+1$.  As a result we
obtain a system which has the form:

\begin{eqnarray}
\dot{a}_i & = & a_i[1-(a_i+\sum_{i \neq j}^{N+M_1}\rho_{ij}a_j)]\label{eq5}
\end{eqnarray}

\noindent with $M_1 \leq M$.  Then we consider the only those indexes
$i$ for which conditions (\ref{eq7}) and (\ref{eq8}) are satisfied. After some
permutation we have the system

\begin{eqnarray}
\dot{a}_i & = & a_i[1-(a_i+\sum_{i \neq j}^{N+M_2}\rho_{ij}a_j)]\label{eq6}
\end{eqnarray}

\noindent with $M_2 \leq M_1$.  Let us form a graph $G$ now: its nodes
are equilibrium points of (\ref{eq6}), e.g.  points $A_i=(0...0 i=1
0...)$. There is an edge starting at the point $A_i$ and ending at the
point $A_j$ if there is a separatrix (a piece of $W_i$) joining and
$A_i$ and $A_j$.  Thus, for any point $A_i$ there is the only one edge
starting at $A_j$.  It implies that there exists a subgraph of this
graph in the form of a cycle that contains, say, $N$ vertices. The
equation describing dynamics of $N$ corresponding neurons has exactly
the form (\ref{eq3}) (maybe after some permutative change of
variables).


\section{Birth of a Stable Limit Cycle}
\label{stable}

A direct corollary of Theorem 1 is the possibility of the birth of the
stable limit cycle in system (\ref{eq3}) when perturbed in an
appropriate way. A perturbation should act in such a way that a
Poincar\'e map $S_{P_1} \rightarrow S_{P_1}$ for the perturbed system
has an absorbing region and a fixed point inside it. Such a condition
can be expressed differently. Let us do it as follows. Consider the
system

\begin{eqnarray}
\dot{a}_i = a_i [1-(a_i+ \sum_{j \neq i}^N \rho_{i j} a_j)] + \epsilon \Psi_i(a)\label{eq25}
\end{eqnarray}

\noindent that coincides with (\ref{eq3}) for $\epsilon=0$, where
$a=(a_1, ..., a_N)$ and $\Psi_i$ is a smooth function,
$i=1,...,N$. For small $\epsilon>0$ the system (\ref{eq25}) has saddle
equilibrium points $A_{i \epsilon}$ and separatrices $\Gamma_{i
\epsilon}$ (the half of $W_{i \epsilon}^u$ such that $A_{i \epsilon}
\rightarrow A_i$, as $\epsilon \rightarrow 0$ and $lt_{\epsilon
\rightarrow 0} \Gamma_{i \epsilon} \supset \Gamma_{i i+1}$ (here $lt$
means the topological limit, i.e. the set of the accumulation points).

\begin{theorem}
Assume that the conditions of Theorem 1 are
satisfied,

\begin{eqnarray}
lt_{\epsilon
\rightarrow 0} (\bigcup_{i=1}^N \Gamma_{i \epsilon}) = \Gamma\label{eq26}
\end{eqnarray}

\noindent and at least one of the separatrices $\Gamma_{i \epsilon}$
is not a heteroclinic orbit. Then for any sufficiently small $\epsilon
>0$ the system (\ref{eq25}) has a stable limit cycle $L_\epsilon$ (in a
neighborhood of $\Gamma$) such that $lt_{\epsilon \rightarrow 0}
L_{\epsilon} = \Gamma$. 
\end{theorem}

The proof of this Theorem can be done in the standard way, i.e., by
construction of the Poincar\'e map and by showing (as in the proof of
Theorem 1) that this map is a contraction in an absorbing region. The
condition (\ref{eq26}) (or a similar condition) is necessary and
sufficient for the existence of an absorbing region. We omit the proof
in the present work, since it is really a standard one (the
corresponding technique can be found in \cite{shilnikov,afraa2,afraa1}.
Thus, one may say that the system under consideration is robust
in the following sense: the attractor of a perturbed system
remains in a small neighborhood of the ``unperturbed'' attractor.



Numerical results show that the system (\ref{eq25}) where $\Psi_i(a)
\geq 0$ satisfies the condition (\ref{eq26}) and has a stable limit
cycle. An example is shown in Fig.~\ref{trayecto}. 
In particular, it is clear from this figure that the system has the 
only one -global- attractor.
In this example,
the simulations were performed with the following equations:

\begin{eqnarray}
\dot{a_i} & = & a_i(1-\sum_{j=1}^{N=6}\rho_{ij}a_j)+\epsilon a_i
a_{i+3} \label{eqarr}
\end{eqnarray}

\noindent where $i=1,2,...,6$ and $i+3 \equiv i-3$ if $i>3$. We used
the following values of the connection matrix $\rho_{ij}\neq 0$:

\begin{eqnarray}
\rho_{1,3}&=&\rho_{3,5}  =  \rho_{5,1}  =  5; \hspace*{2mm} \rho_{4,6} = \rho_{2,4}  =  \rho_{6,2}  =  2\nonumber \\
\rho_{1,6}&=&\rho_{2,1}  =  \rho_{3,2}  =  \rho_{4,3}  =  \rho_{5,4}
=  \rho_{6,5}  =  1.5\nonumber\\
\rho_{1,1}&=&\rho_{2,2}  =  \rho_{3,3}  =  \rho_{4,4}  =  \rho_{5,5}  =  \rho_{6,6}  =  1
\label{ros2}
\end{eqnarray}
\noindent with $\epsilon=0.01$.

Time series showing the switching of activities displayed in this
system by the WLC are depicted in Fig.~\ref{switch}.

\section{WLC in Real Neural Systems}
\label{realn}

In this section we describe two examples of living neural systems that
appear to use WLC strategies to process sensory information. As we
discussed in the Introduction, sensory systems need both robustness
and sensitivity to acquire and process incoming signals. The WLC among
sensory neurons guarantees the coexistence of these essential
characteristics and can explain the experimental recordings in several 
sensory systems. We will illustrate two of these sensory systems using
models with WLC dynamics.

\subsection{WLC in Olfactory Processing}
\label{olfa}
Observed features of olfactory processing networks
\cite{wehr,annrev} can guide the study of computation using
competitive networks. In Fig.~\ref{experim} we show the simultaneously
recorded  activity of three different projection neurons (PNs) in
the locust olfactory system, i.e. antennal lobe (AL), evoked by two
different odors:
despite similar PN activities before the stimulus onset (the result
of the action of noise) each odor evokes a specific spatio-temporal
activity pattern that results from interactions between these and
other neurons in the network~\cite{annrev,bazhen,jphys}.

\begin{figure}[hbt]
  \begin{center}
\leavevmode
    \epsfxsize=5.5 cm
    \epsffile{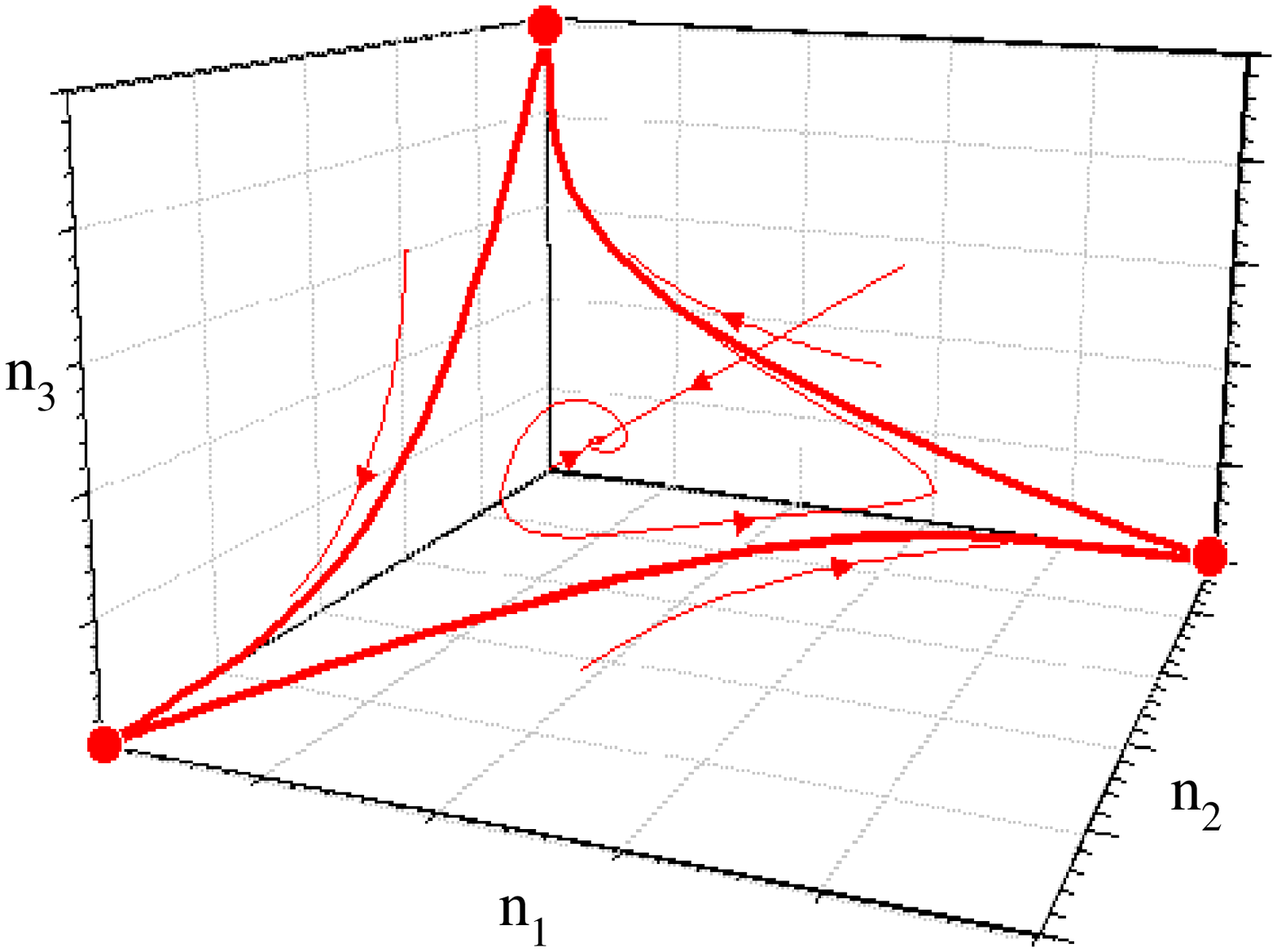} \hspace{5mm}
    \epsfxsize=6 cm
\epsffile{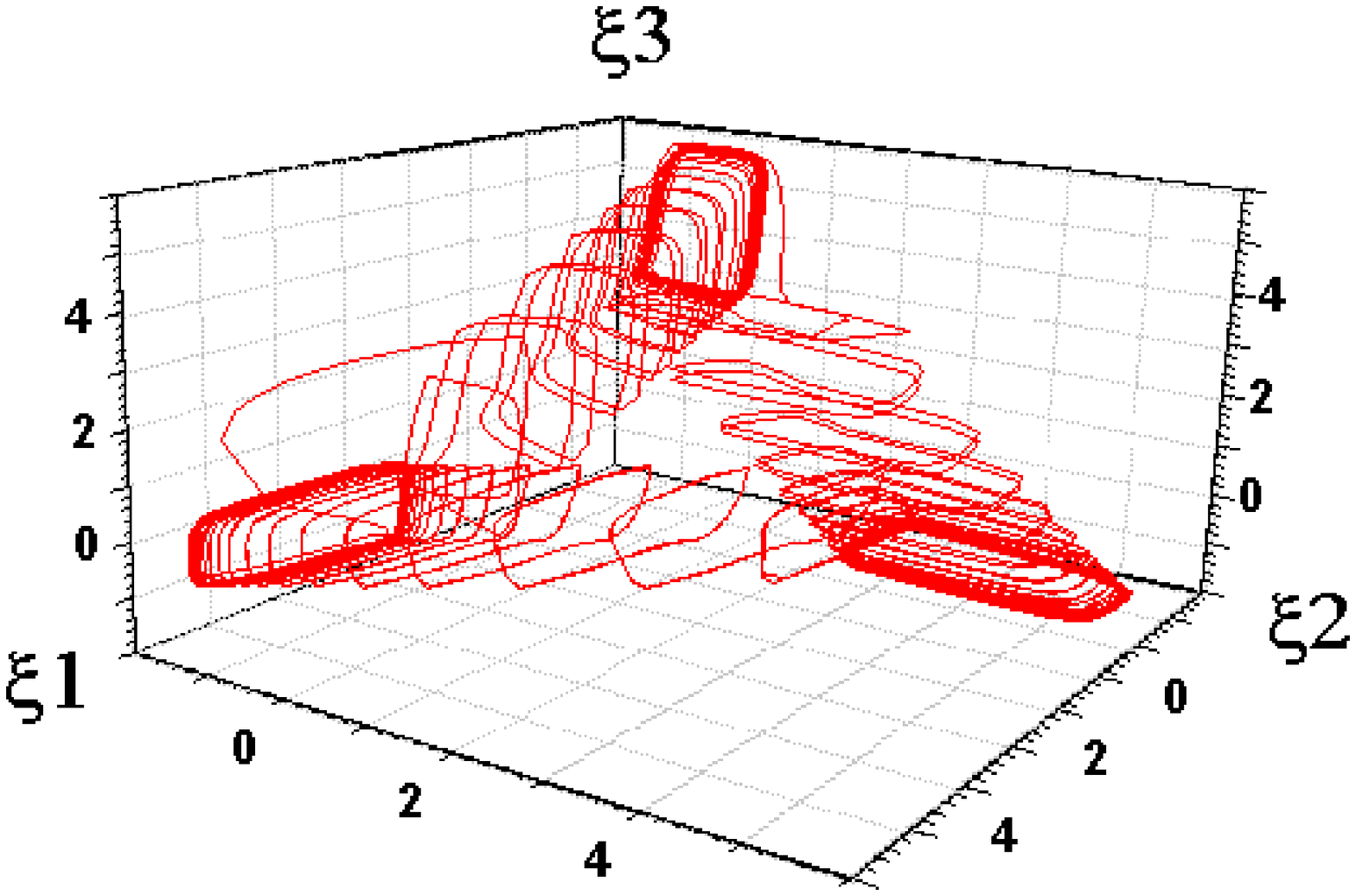}
  \end{center}
\vspace*{-1cm}
\caption{Left panel: phase protrait corresponding to the autonomous WLC dynamics
of a three-dimensional case. Right panel: projection of a
nine-dimensional heteroclinic orbit of three inhibitory coupled
FitzHugh-Nagumo spiking neurons in a three
dimensional space (the variables $\xi_1$, $\xi_3$, $\xi_3$ are
linear combinations of the actual phase variables of the
system~\protect\cite{jphys}) --see also Sec.~\ref{olfa}--.}
\label{Fig1}
\end{figure}

\begin{figure}[ht!]
\begin{center}
\epsfxsize=11cm
\epsffile{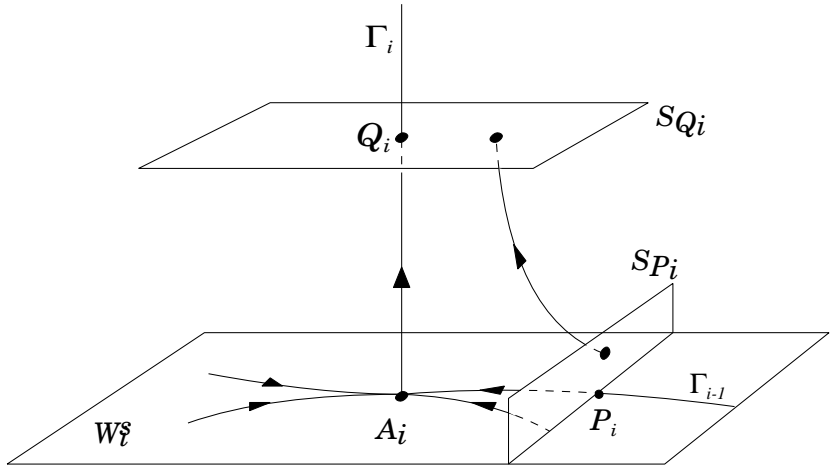}
\end{center}
\caption{\small The local map in a neighborhood of a saddle point.}
\label{trayectof} 
\end{figure}

\begin{figure}[ht!]
\begin{center}
\epsfxsize=11cm
\epsffile{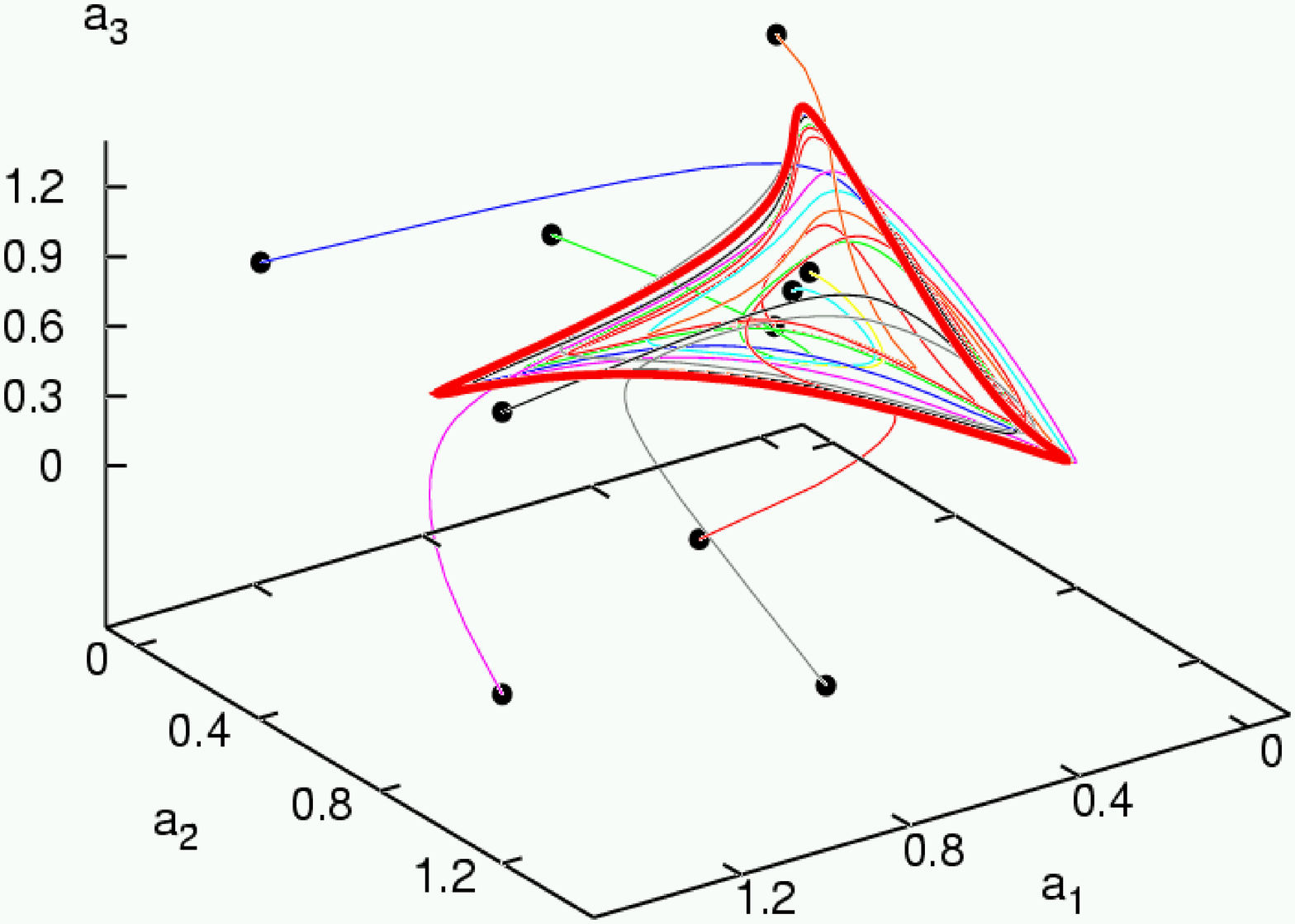}
\end{center}
\vspace*{-1cm}
\caption{\small 3D projection of the 6-dimensional system (\ref{eqarr})
showing  examples of trayectories falling into the limit cycle from
different initial conditions. As the numerical results indicate, this
limit cycle in the vicinity of the former heteroclinic contour is a
global attractor}
\label{trayecto} 
\end{figure}

\begin{figure}[ht]
  \begin{center}
\leavevmode
    \epsfxsize=10cm
    \epsffile{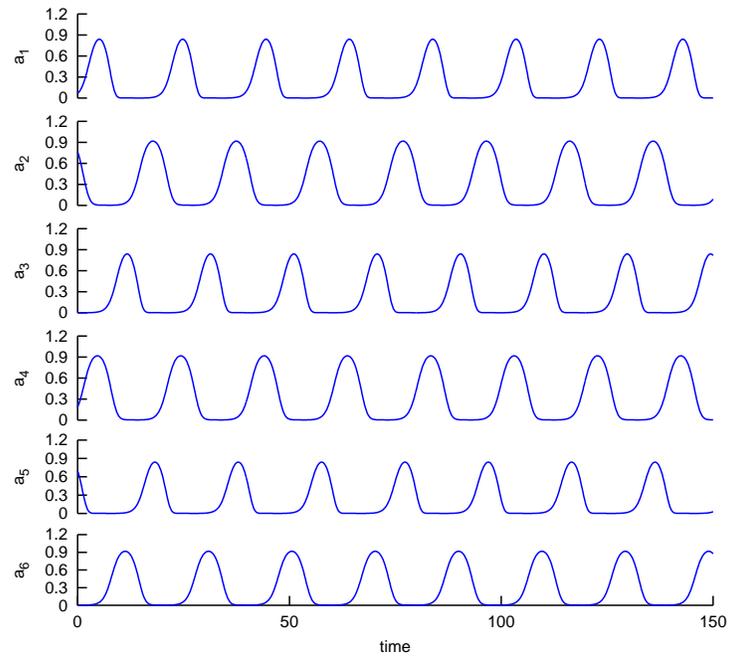}
  \end{center}
\vspace*{-0.5cm}
\caption{\small Time series showing the  switching of activities
$a_i$ in a network of six neurons described by
  equations (\ref{eqarr}). Units are dimensionless. See parameters used in the text.}
\label{switch} 
\end{figure}

\begin{figure}[hbt]
  \begin{center}
\leavevmode
    \epsfxsize=7cm
    \epsffile{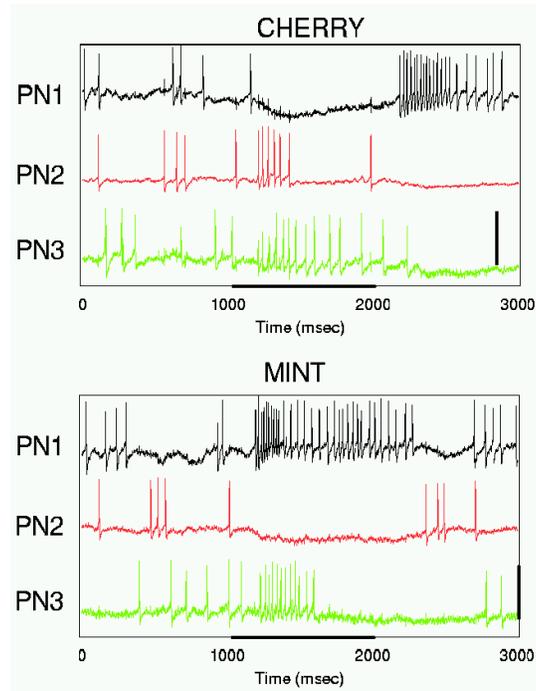} \hspace{5mm}
  \end{center}
\hspace*{-1cm}
\vspace*{-1cm}
\caption{Temporal patterns produced by three
simultaneously sampled PNs in the locust antennal lobe when two
different odors are presented during the time interval from 1000 to 2000
msec. The horizontal bar indicates the
time interval when the stimulus was presented
(see~\protect\cite{wehr} for details).}
\label{experim}
\end{figure}

As we have discussed, WLC networks produce identity-temporal or
spatio-temporal coding in the form of deterministic trajectories
moving along heteroclinic orbits that connect saddle fixed points or
saddle limit cycles (see Fig.~\ref{Fig1}) in the system's state
space. These saddle states correspond to the activity of specific
neurons or groups of neurons and the separatrices connecting these
states correspond to sequential switching from one state to another.


From the experimental results~\cite{wehr,annrev} we infer that a
stimulus acts in two principal ways: (1) it excites a subset of
projector neurons; (2) it modifies the effective inhibitory
connections between the projector neurons as a result of activation of
the inhibitory interneurons that connect different PNs. The intrinsic
dynamics of these neurons is governed by many variables corresponding
to ion channels and intracellular processes. Such detailed description
however is not needed to illustrate the principle of ``coding with
separatrices''. We need only to capture the `firing' or `not-firing'
state of the component neurons. We thus simplify our model to an
equation for the firing rate $a_i(t) > 0$ of neural
activity~\cite{jphys}:
\begin{equation}
\dot{a}_i(t)= a_i(t) \bigg[\sigma_i({\bf S})-\bigg( a_i +
\sum_{j \neq i}^N \rho_{ij}({\bf S})a_j(t)\bigg)\bigg]+S_i(t).
\label{eqdos}
\end{equation}
\noindent where $\rho_{ij}({\bf S})$ is the strength of inhibition of
neuron $j$ onto $i$. $\sigma_i({\bf S})=-1$, when there is no
stimulus, and $\sigma({\bf S})=+1$ when the stimulus has a component
at neuron $i$. When $\sigma_i = -1$, the quiet resting state $a_i =
0$ is stable. When a stimulus is applied and $\sigma_i = +1$, the
system moves away from this quiet state onto a sequence of
heteroclinic trajectories. This instability triggers the system into
rapid action, provides robustness against noise and allows a response
independent of the state at stimulus onset.

 When the inhibitory connections are not symmetric, the system with $N$
competitive neurons has different closed heteroclinic orbits that
consist of saddle points and one dimensional separatrices connecting
them.  Such heteroclinic orbits are global attractors in phase space
and are found in various regimes of the $\rho_{ij}({\bf S})$. This
implies that if the stimulus is changed, another orbit in the vicinity
of the heteroclinic orbit becomes a global attractor for this
stimulus. Such rich behavior can be illustrated also by an inhibitory
ensemble of spiking neurons. We have studied a network of inhibitory
connected FitzHugh-Nagumo neurons  ($i = 1, 2, ... , 9$)~\cite{jphys}:
\begin{eqnarray}
\tau_1 \frac{dx_i(t)}{ dt} & =
&f(x_i(t))-y_i(t)-z_i(t)(x_i(t)-\nu)+0.35+S_{i}   \nonumber \\
\frac{ dy_i(t)}{dt} & = & x_i(t)-by_i(t)+a \nonumber \\
\tau_2 \frac{ dz_i(t)}{ dt} & = & \sum_j  g_{ji}G(x_j(t))-z_i(t)
\label{FH}
\end{eqnarray}
Here we use a dynamical model of inhibition: $z_i(t)$ is a synaptic
current modeled by first order kinetics. The variable $x_i(t)$
denotes the membrane potential, $y_i(t)$ is a recovery variable,
and $f(x)=x-\frac{1}{3}x^{3}$ is the internal FN nonlinearity. The
stimulus is taken as a constant. We use a step function for $G(x) =
0, x \leq 0, \mbox{and} \, G(x) = 1, x > 0$, as the synaptic
connection. $S_{i}$ is the stimulus, and $g_{ji}$, the strength of
synaptic inhibition: $g_{ji}=2$ if the $j^{th}$ neuron inhibits the
$i^{th}$; $0$ otherwise. The other parameters are $a=0.7,
b=0.8,\tau_{1}=0.08,\tau_{2}=3.1, \nu=-1.5$.

\begin{figure}[ht]
  \begin{center}
\leavevmode
    \epsfxsize=7 cm
    \epsffile{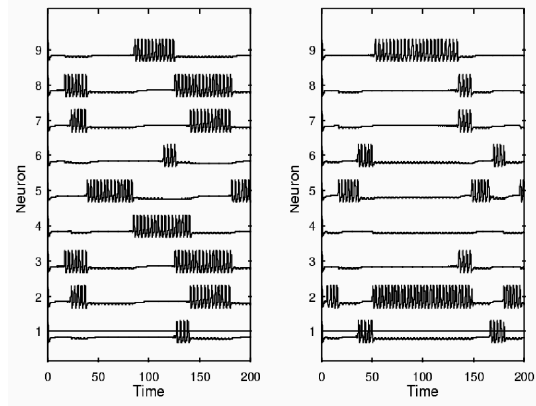}
  \end{center}
\caption{Spatio-temporal patterns generated by a network of nine
FitzHugh-Nagumo neurons with inhibitory connections. We used the
external stimuli: $S_{1}=0.1$, $S_{2}=0.15$, $S_{3}=0$, $S_{4}=0$,
$S_{5}=0.15$, $S_{6}=0.1$, $S_{7}=0$, $S_{8}=0$, $S_{9}=0$,
$\tau_{2}=3.1$ (left), and $S_{1}=0.01$, $S_{2}=0.03$,
$S_{3}=0.05$, $S_{4}=0.04$, $S_{5}=0.06$, $S_{6}=0.02$,
$S_{7}=0.03$, $S_{8}=0.05$, $S_{9}=0.04$, $\tau_{2}=4.1$ (right).
We plot $x_i(t)$ versus time.
\label{Traces}
}
\end{figure}

Our numerical simulations show that the network produces different
spatio-temporal patterns in response to different
stimuli. Fig. \ref{Traces} presents examples of these activities
corresponding to two different stimuli. The system was in the resting
state $x_{i}\approx -1.2, y_{i}\approx -0.62, z_{i}=0$ before the
stimulus began at $t=0$. As one can see, the patterns are considerably
different and distinguishable. The heteroclinic contour in this
network consists of a finite number of saddle limit cycles and the
same number of heteroclinic orbits connecting these cycles (see, as an
example, Fig.~\ref{Fig1}, right panel). A detailed characterization of
this network as an information processing device has been reported
in~\cite{mishaprl}.

\subsection{WLC in the Gravimetric Neurons of the Mollusk
{\sl \bf Clione}}

\begin{figure}[ht]
  \begin{center}
\leavevmode
    \epsfxsize=6cm
    \epsffile{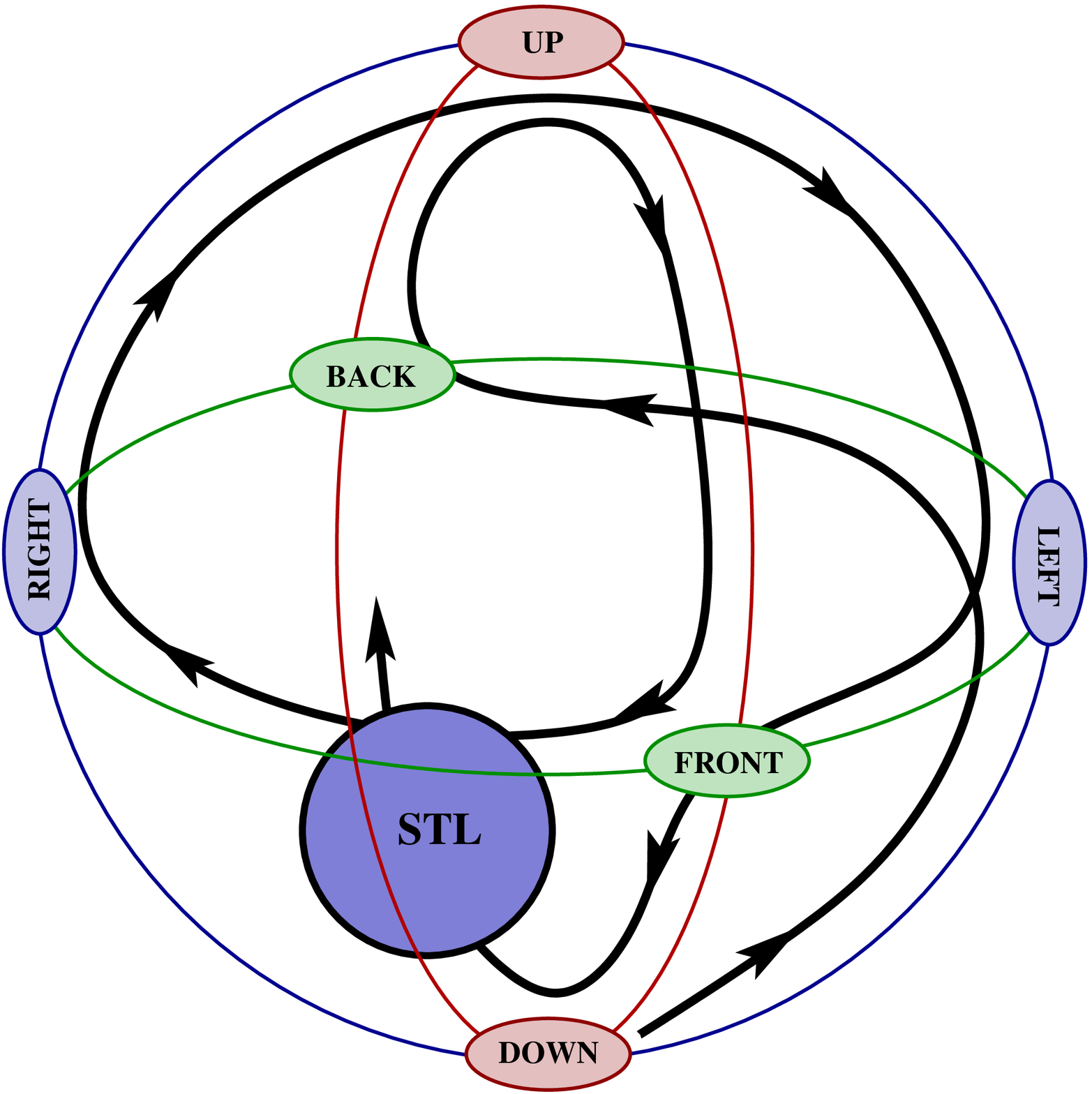}
    \hspace*{1cm}
    \epsfxsize=5.5cm
    \epsffile{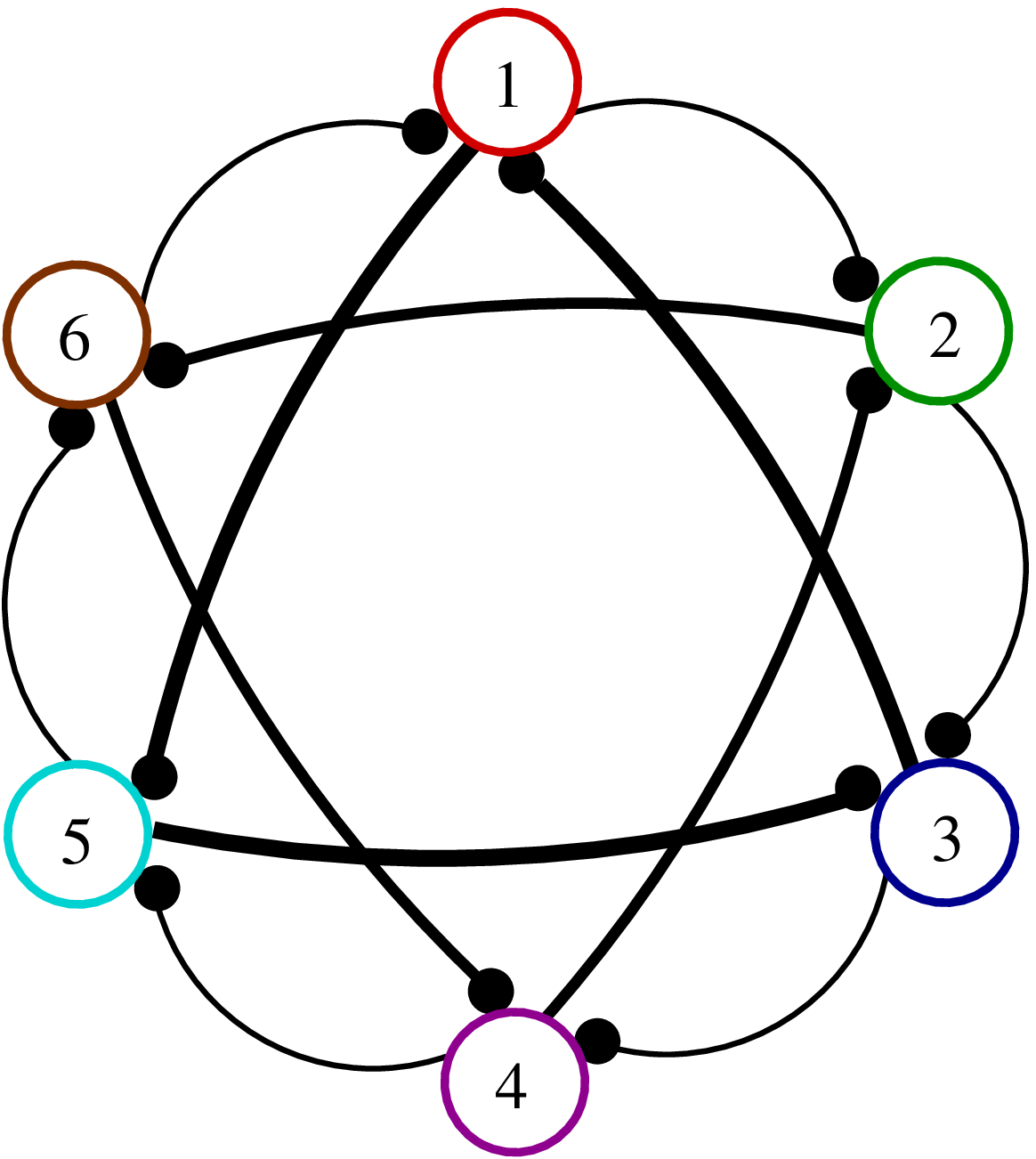}
  \end{center}
\caption{Left panel: Schematic representation of the statolith motion
inside the statocyst, the gravimetric organ of the mollusk {\sl
Clione}. Right panel: inhibitory connections used in the network model
of the statocyst receptor neurons (thicker traces mean stronger
inhibition).}
\label{statocyst}
\end{figure}

Neural networks with WLC dynamics are able to generate new information
to answer a simple external signal. Such information can be used for
the organization of complex activity and, in particular, chaotic
behavior of some animals. Let us consider now the hunting activity of a
marine mollusk {\sl Clione}.  This mollusk is a predator lacking a
visual system. It feeds on a small mollusk, {\sl Limacina}.  The
hunting behavior is a random search for prey: {\sl Clione} "scans" the
surrounding space in order to locate and catch the prey. Such behavior
is turned on by the smell of the {\sl Limacina}. The main role in the
organization of such motion of {\sl Clione} is played by a sensory
neural network inside the gravimetric organs: the statocysts (see
Fig.\ref{statocyst}).  These special sensory organs are 
responsible for the orientation in the gravitational
field~\cite{yura}.



It is well known from the physiological data that the statocysts have
up to 12 receptor neurons (SRNs) that are coupled with inhibitory
synapses~\cite{yura}. These neurons respond to the pressure exerted by
the statolith, a stone located inside the statocyst. If no information
about a prey (received by the chemical receptors) is present, the
receptor neuron $D$ (down, see Fig.~\ref{statocyst}) is excited by the
statolith and it inhibits other SRNs, and the network responds in a
winner-take-all mode. As a result, the information generated by $D$ SRN
arrives to the corresponding Central Pattern Generators (CPGs) that
control the tail and wing movements. These CPGs establish the habitual
"head up" position of {\sl Clione}'s body. However if a special
Hunting Central Neuron (HCN) receives a message from the chemo-sensors
about the presence of a prey, HCN excites the SRNs organizing a WLC
among them as we will illustrate with a model. The behavior of the
{\sl Clione} in this case does not depend on the direction of the
gravitational field and it moves in a random-like trajectory.

      For the phenomenological modeling of the statocyst ``hunting''
dynamics we can neglect the statolith inertial dynamics and take into
account the only key point: the position of the mollusk's body
uniquely depends on the message that SRNs are sending to the central
neurons that produce the commands to the CPGs. Thus, as a starting
point, we consider just a SRN network under the action of the HCN
excitation. We suppose that, as a result of the HCN stimulation, all
SRNs ("left", "right", "back", "front", "down", and "up") are in the
same situation: they receive and send two inhibitory synapses (see
Fig.~\ref{statocyst}, right panel).

The dynamics of the SRN's network can be described by
model (\ref{eq2}) with $N=6$. In this case, $a_i>0$ represents the
instantaneous spiking rate of the receptor neuron $i$, $H_i(t)$
represents the stimulus from the hunting neuron to neuron $i$, and
$S_i(t)$ represents the action of the statolith on the receptor that
is pressing. When there is no stimulus from the hunting neuron
($H_i=0, \forall i$) or the statolith ($S_i=0, \forall i$), then
$\sigma(\mbox{\boldmath $H$},\mbox{\boldmath $S$})=-1$ and all neurons
are silent; $\sigma(\mbox{\boldmath $H$},\mbox{\boldmath $S$})=1$ when
the hunting neuron is active and/or the statolith is pressing one of
the receptors. In our simulations, we have used the values $\rho_{ij}
\neq 0$ specified in (\ref{ros2}).

When there is no activation of the sensory neurons from the hunting
neuron, the effect of the statolith ($S_i \neq 0$) in this model is to
induce a higher rate of activity on one of the neurons (the neuron $i$
where it rests for a big enough $S_i$ value).  We assume that this
higher rate of activity affects the behavior of the motoneurons to
organize the head up position. The other neurons are either silent or
have a lower rate of activity and we can suppose that they do not
influence the posture of {\sl Clione}.

\begin{figure}[ht]
  \begin{center}
\leavevmode
    \epsfxsize=7.2cm
    \epsffile{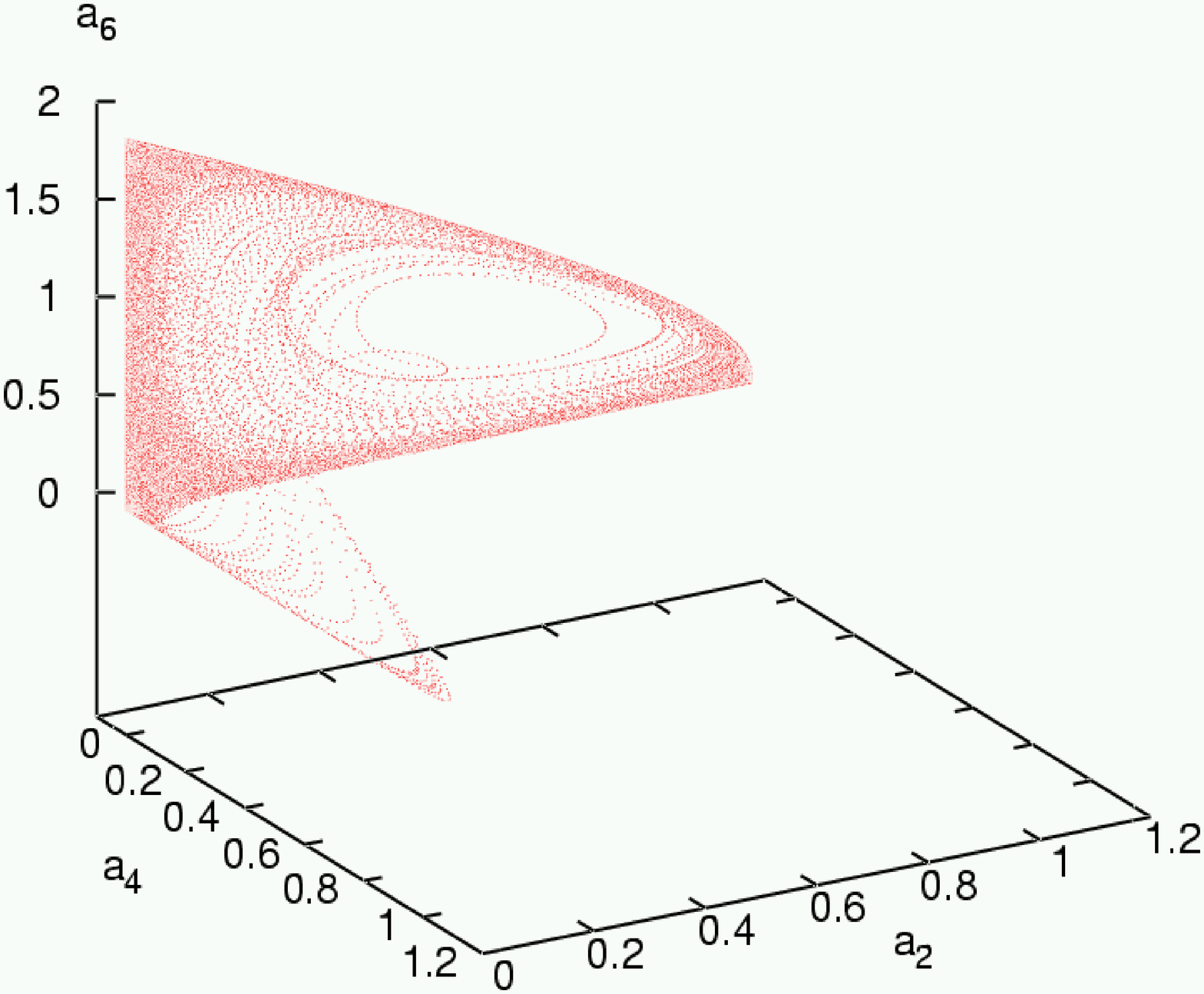}
    \hspace*{-1cm}
    \epsfxsize=7.2cm
\epsffile{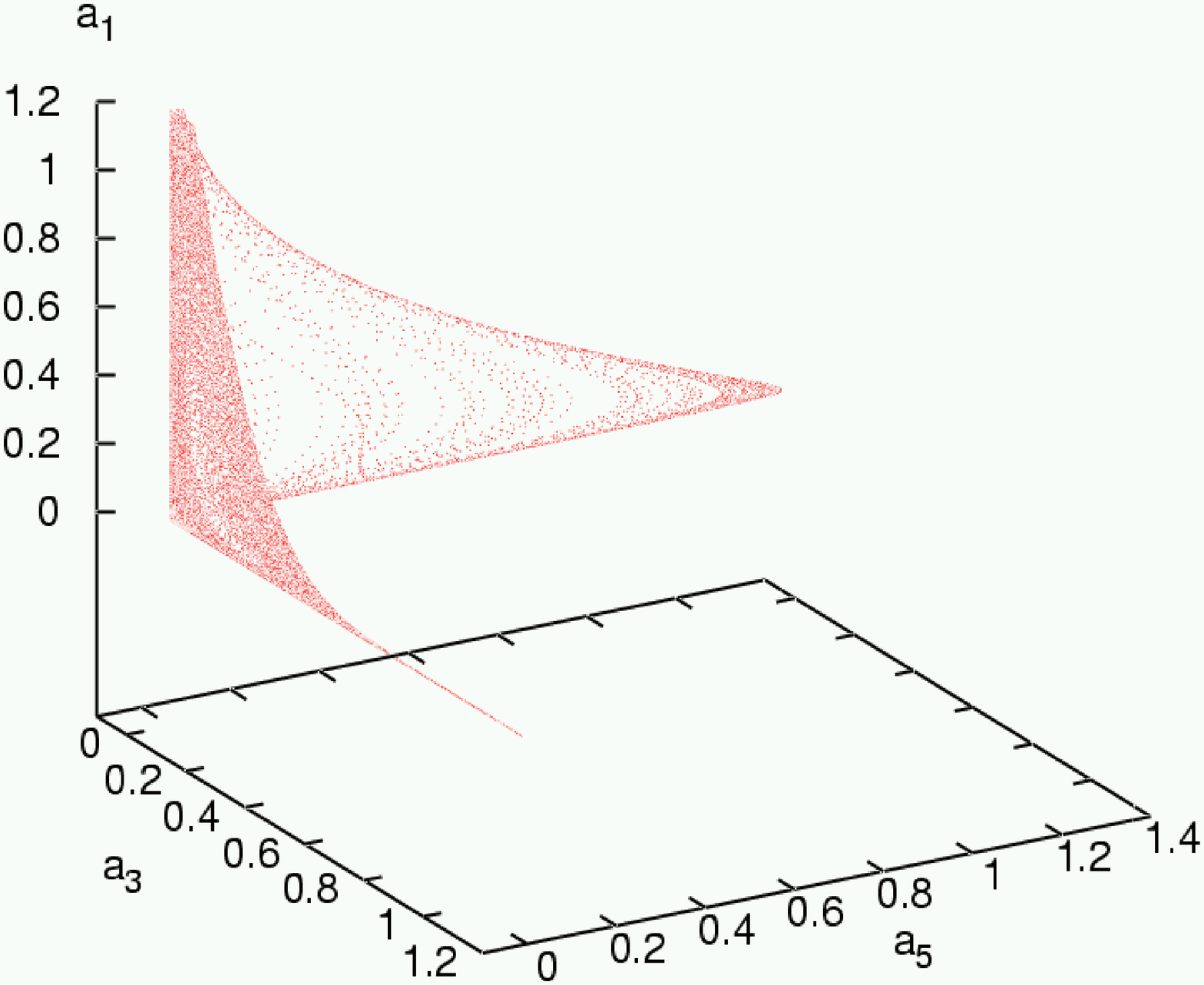}
  \end{center}
\caption{Projections of the attractor from the six-dimenional phase
space of the statocyst receptor network to two different three-dimensional spaces.}
\label{clioneat}
\end{figure}

When the hunting neuron is active a completely different behavior
arises. We assume that the action of the hunting neuron overrides the
effect of the statolith and thus $S_i\approx 0, \forall i$.  The
dynamical system (\ref{eq2}) with the $\rho_{ij}$ values specified
above (see also Fig.~\ref{statocyst}) and with a stimuli from the
hunting neuron given, for example, by
$H_i=(0.730,0.123,0.301,0.203,0.458,0.903)$ has a strange attractor in
the phase space (see Fig.~\ref{clioneat}). This means that the SRN
network generates new information (a chaotic signal with positive
Kolmogorov-Sinai entropy) in the presence of the prey, which controls
the CPGs and, in fact, organizes the random-like behavior of {\sl
Clione}.

      The origin of the chaoticity in such dynamical system can be
explained in the following manner~\cite{clione}: due to the diversity
in the strengths of the inhibitory connections we may consider the
complete network as two weakly coupled WLC triangle
networks. Independently each of them has a closed heteroclinic contour 
(see Fig.~\ref{statocyst}), which becomes a limit cycle under the action of a
small perturbation (see Sec.~\ref{stable}). The periodic
oscillations corresponding to these limit cycles have, in general,
different frequencies that are extremely sensitive to the distance to
the heteroclinic loop in the non perturbed system (such oscillations
are strongly non-synchronous). As we showed the weak interaction of
these WLC triangles (nonlinear oscillators) generate chaos in wide
regions of the control parameter space. New experiments have confirmed
the validity of the model and its predictions~\cite{sfn02}.

\newpage

\section{Discussion}

The stimulus dependent sequential switching of neurons or group of
neurons (clusters), named WinnerLess Competition, is able to solve the
fundamental contradiction between sensitivity and robustness of the
sensory recognition. The key points on which the WLC networks are
based are: (i) the heteroclinic contour corresponding to a specific
sequence of switching has a large basin of attraction, i.e. a specific
sequence is stable; and (ii) the topology of the heteroclinic contour
sensitively depends on the incoming signals, i.e. high resolution or
sensitivity. Both features are actually present only in the case if
under the action of a perturbation the discussed heteroclinic contour,
which is structurally unstable, is transformed to the limit cycle in
its vicinity with the same topology as the contour. In this paper we
have discussed the conditions for such topological stability, and we
have showed that computing with separatrices based on the WLC
principle is a very natural and powerful strategy for information
processing in real neural systems. Any kind of sequential activity can
be programmed by a network with stimulus dependent nonsymmetric
inhibitory connections. It can be the creation of spatio-temporal
patterns of motor activity, the transformation of the spatial
information into spatio temporal information for successful
recognition and many other computations. In addition, we wish to
mention that two important computational functions can be successfully
implemented by computation with separatrices. These are: (i)
sequential memory storage, and (ii) feature binding.

In reference~\cite{seliger} the authors suggest a new biologically-motivated
model of sequential spatial memory which is based on the WLC
principle. Each stimulus event (visual image, odor, etc...) is represented by
a saddle point in the phase space of the system, and a network of
one-dimensional separatrices leads the system along the sequence of
events in the specific episode. After the learning process, such
system is capable of an associative retrieval of the pre-recorded sequence
of spatial patterns. 

A binding problem occurs when two (or more) different events,
e.g. scenes, features, or behaviors are represented by different
neural ensembles simultaneously, and for some reason they are all
connected with each other. Eventually, these coherent features are
integrated by the nervous system of the animal onto a perceptual
object, even if the features are dispersed among different sensory
systems or subsystems.  The binding is ubiquitous and occurs whenever
a simultaneous remembrance or representation is important. The most
common approach in the modeling of binding is to involve time in
operation (von der Malsburg, Singer, and others). The idea is to use
the coincidence of certain events in the dynamics of different neural
units for binding. This is a dynamic binding. Usually, dynamic binding
is represented by synchronous neurons or neurons that are in resonance
with and external field. However, dynamical events like phase or
frequency variations usually are not very reproducible and robust. It
is reasonable to hypothesize that brain circuits that display
sequential switching of neural activity~\cite{abeles} use the
coincidence of this switching to implement dynamic binding of
different WLC networks.

In the conclusion we have to emphasize that for large inhibitory neural
ensembles it is not necessary to have specific connections that
satisfy the conditions formulated above for the existence and
stability of the WLC dynamics. If the connections are random, the
neurons in the ensemble can be separated in three groups: (i)
neurons that are weakly coupled with others (they behave like nearly
independent elements), (ii) neurons with strong but close to
symmetric connections (they form a Hopfield like network just with
simple attractors), and (iii) neurons with nonsymmetric connections
that demonstrate the WLC dynamics. Because the WLC dynamics is a
structurally stable phenomenon, it is reasonable to hypothesize that
the perturbation of the third group by the first two does not destroy
the switching activity. Recent computer experiments~\cite{zhigulin}
have confirmed this hypothesis.

\newpage
\begin{center}
{\bf ACKNOWLEDGMENTS}
\end{center}

Support for this work came
from NIH grant 2R01 NS38022-05A1, Department of Energy grant
DE-FG03-96ER14592 and NSF/EIA-0130708. V.A. was supported by CONACyT
grant 485100-3-36445-E and by UC MEXUS-CONACyT grant. P.V. was
supported by MCyT BFI2000-0157.

\newpage 


\newpage 


\newpage

\newpage 

\newpage 

\newpage

\newpage

\newpage

\newpage

\end{document}